\RequirePackage{fix-cm}
\documentclass[smallextended]{svjour3}       
\smartqed  
\usepackage{graphicx}
\usepackage{tcolorbox}
\begin{document}


\title{SIEVE: Helping Developers Sift Wheat from Chaff via Cross-Platform Analysis}



\author{Agus Sulistya         \and
		Gede Artha Azriadi Prana \and	
        Abhishek Sharma	\and
        David Lo \and
        Christoph Treude
}


\institute{Agus Sulistya, Gede Artha Azriadi Prana, Abhishek Sharma and David Lo\at
              Singapore Management University \\
              \email{\{aguss.2014,arthaprana.2016,abhisheksh.2014,davidlo\}@smu.edu.sg}           
           \and
           Christoph Treude \at
              University of Adelaide \\
              \email{christoph.treude@adelaide.edu.au}
}

\date{Received: date / Accepted: date}

\maketitle
\sloppy
\begin{abstract}
Software developers have benefited from various sources of knowledge such as forums, question-and-answer sites, and social media platforms to help them in various tasks. Extracting software-related knowledge from different platforms involves many challenges. 
In this paper, we propose an approach to improve the effectiveness of knowledge extraction tasks by performing cross-platform analysis. Our approach is based on transfer representation learning and word embeddings, leveraging information extracted from a source platform which contains rich domain-related content. The information extracted is then used  to solve tasks in another platform (considered as target platform) with less domain-related contents. We first build a word embeddings model as a representation learned from the source platform, and use the model to improve the performance of knowledge extraction tasks in the target platform. We experiment with Software Engineering Stack Exchange and Stack Overflow as source platforms, and two different target platforms, i.e., Twitter and YouTube. Our experiments show that our approach improves performance of existing work for the tasks of identifying software-related tweets and helpful YouTube comments.
\end{abstract}

\maketitle

\section{Introduction}

Software developers rely on many sources of knowledge to help them keep up-to-date on the latest technologies and to solve their problems. A study conducted by Maalej et al.~\cite{maalej2014comprehension} shows that 70 percent of developers use online resources (web search engine, public documentation, forums) as channels to access knowledge. Among those channels, some are more popular and contain richer content relevant to software engineering compared to others. For example, Xin et al.~\cite{xia2017developers} conducted an observation on how developers commonly make use of a web search engine such as Google to find online resources to improve their productivity. They found that 63\% of the searches on the Internet ended up with a visit to Stack Overflow, a popular question and answer (Q\&A) site.

Despite the popularity of Stack Overflow, software developers also seek knowledge in other platforms, such as  microblogging platforms (e.g., Twitter). A large number of software developers  use Twitter frequently to support their professional activities, e.g., to share and obtain the latest technical news~\cite{singer2014software}. Another growing popular knowledge source for developers is online video sharing platforms, such as YouTube. A study by MacLeod et al.~\cite{macleod2015code} found that video is a useful medium for communicating knowledge between developers, and that developers build their online personas and reputation by sharing videos through social channels. Developers as content creators will also be able to digest feedback from valuable comments given by their viewers. It will help content creators to improve their future videos~\cite{poche2017analyzing}.

Extracting software-related knowledge from different platforms requires varying levels of effort and skill. For example, on Stack Overflow, almost all of the contents are related to software development. The content is also maintained to be of high quality by collective community effort and the site's moderators. But it is more challenging to extract useful information from tweets, due to the information overload problem in Twitter's space. Twitter is a popular social media channel with about 330 million users who produce about 500 million tweets daily~\cite{abouttwitter2}. Singer et al. found that Twitter is popular among software developers also, who use it to \textit{keep up with the fast-paced development landscape}~\cite{singer2014software}. Around 70\%  of the respondents said that they use Twitter to stay current about the latest technologies, practices, and tools, and  also to learn about things that they are not actively looking for. A number of them also said that they use Twitter for community building especially around their development projects.  They also found that developers face challenges while using Twitter, which relate to dealing with a huge amount of irrelevant tweets produced on Twitter, as well as the challenge of maintaining a relevant network. With a huge amount of content being produced by a large number of users, developers face a hard time in finding tweets with information relevant to software development. Twitter is a channel which is very noisy with information and users from domains other than software engineering. This gives rise to the problem of information overload for software developers who use Twitter. 


In this paper, we propose \texttt{SIEVE}, an approach to utilize contents from a rich software-development-specific platform to help automated knowledge extraction tasks in other less software-development-specific platforms, based on a transfer representation learning approach. We consider two platforms, Software Engineering Stack Exchange and Stack Overflow, as the rich domain-related platforms. We build word embeddings based on the dataset collected from these platforms. We then leverage the word embeddings vectors to solve information retrieval and classification problems in two different target platforms. We experiment with two different use cases: finding tweets relevant to software development on Twitter~\cite{sharma2015nirmal}, and classifying useful comments for software engineering video tutorials on YouTube~\cite{poche2017analyzing}. We conducted experiments based on the existing golden datasets provided by Sharma et al.~\cite{sharma2015nirmal} for Twitter, and Poch\'e et al.~\cite{poche2017analyzing} for YouTube comments. Our experiments show the effectiveness of our proposed cross-platform analysis approach which achieves performance improvements of up to 23\%  and 10.3\% for the first and second use case respectively. Our contributions can be summarized as follows:

\begin{enumerate}
	\item We propose an approach based on transfer representation learning and word embeddings to solve information retrieval problems on how to use data from domain-specific platforms to help tasks in other platforms.
	\item We conduct experiments to show the effectiveness of the proposed approach for two different tasks and platforms (i.e., Twitter and YouTube), and use baselines described in existing work.
\end{enumerate}

The next sections in this paper are structured as follows. In Section~\ref{sec.background}, we describe background related to knowledge channels for software developers, and background on representation learning and word embeddings. In Section~\ref{sec.approach}, we describe our approach on learning a knowledge representation from source platforms. We present our first use case on finding software-related tweets in Section~\ref{sec.task1}. Next, we present the second use case on classifying informative comments on YouTube in Section~\ref{sec.task2}. Threats to validity are discussed in Section~\ref{sec.threats}. We describe related work in Section~\ref{sec.related}. Finally, we conclude and mention future work in Section~\ref{sec.conclusion}. 
\section{Background}
\label{sec.background}
In this section, we first discuss the knowledge sources used by developers which we have considered in our current work. Next, we discuss some background on transfer representation learning and word embeddings.

\subsection{Knowledge Sources for Software Developers}

Storey et al. found that software developers use many communication tools and channels in their software development work~\cite{storey2014r,storey2017social}. In our current work we focus on learning word embeddings from software-development-specific channels such as Software Engineering Stack Exchange and Stack Overflow (which are popular software discussion forums), and use the learned embeddings to improve the performance of information retrieval and classification tasks related to the extraction of software-development-related knowledge from open domain channels such as Twitter (a microblogging site) and YouTube (video sharing). In the subsequent paragraphs we give background on these channels. 

\vspace{0.2cm}\noindent{\textit {Software Engineering Stack Exchange}}: Stack Exchange\footnote{https://stackexchange.com/} is a network of question and answer (Q\&A) websites, where each website focuses on a specific topic. On any of the websites each of which is related to a particular domain, its users can ask questions related to that domain and other users can provide answers to these questions. The motivation for users to answer questions comes from the points that they can gain when other users in the same community upvote or accept their answers. These points help them to build a reputation in the domain (and the related community), which the Stack Exchange website is focused on. The Stack Exchange community has been the focus of many studies such as~\cite{begel2013social,posnett2012mining}. In this work as we are interested in improving the  performance of information retrieval  and classification tasks related to  software engineering, we focused on Stack Exchange communities focused on software engineering and programming, which are Software Engineering Stack Exchange\footnote{https://softwareengineering.stackexchange.com/} and Stack Overflow{\footnote{https://stackoverflow.com/\label{stackoverflowurl}} respectively. The difference between these two sites is that 
\textit{Stack Overflow} is focused only on specific programming tasks and problems, whereas \textit{Software Engineering Stack Exchange} allows more general questions related to software development and engineering such as discussions about various libraries, methodologies etc. The latter has about 50,655 questions and 260,361 users. The  intuition behind using Software Engineering Stack Exchange is that models trained on the  general nature of content may achieve different performance on the task of  filtering information from open domain websites such as Twitter and YouTube. 

\vspace{0.2cm}\noindent{\textit {Stack Overflow}}: 
Stack Overflow\textsuperscript{\ref{stackoverflowurl}} is a programming question and answer website founded in 2008 with a focus on software development. It is an online forum where anybody facing a programming issue can post a question describing the problem they face. The questions posted are public on the forum, so any other user on the forum can post their solutions as answers to the posted questions. The original asker can then mark an answer as accepted if it solved the problem. Other users can also upvote an answer if they think it is the right method to solve the programming challenge being addressed. Thus Stack Overflow helps developers in getting answers to their problems with the help of the crowd. It is one of the most used websites by software developers in the world having more than 9,000,000  registered users, more than 16,000,000 questions and an Alexa Rank of 70\footnote{https://en.wikipedia.org/wiki/Stack\_Overflow}. As Stack Overflow contains rich software development and software engineering content, it has been immensely popular among  software engineering researchers in recent years, where it has been used to discover topics and trends~\cite{barua2014developers}, generate API call rules~\cite{azad2017generating}, explore knowledge networks~\cite{ye2017structure}, build information filtering models~\cite
{sharma2015nirmal} etc. More related work is discussed in detail in  Section~\ref{sec.related}.

\begin{figure*}[!htb]
	\centering
	\scalebox{0.65}{\includegraphics{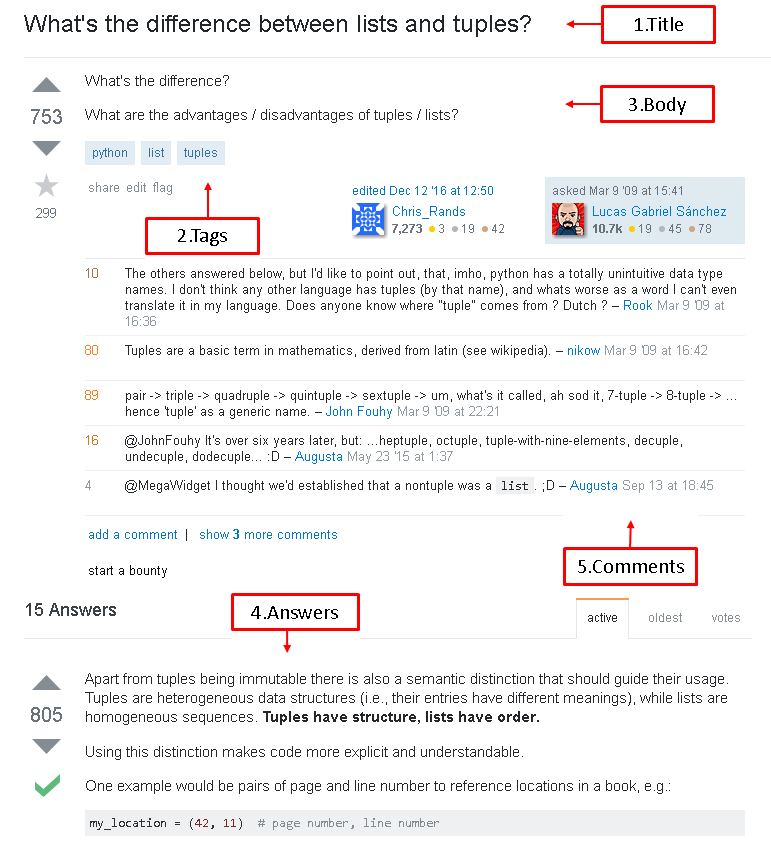}}
	\caption{A Sample question-answer-thread on Stack Overflow with tags (Thread ID 626759)}\label{sec.approach.figso1}
	\label{figso1}       
\end{figure*}

Figure~\ref{sec.approach.figso1} shows a sample question-and-answer thread from Stack Overflow. Each thread generally contains five types of information: title, tags, body, answers, and comments. The title of a thread is a summary of the question asked. The tags represent  the metadata related to the question being asked and are entered by the person who asked the question. Whenever somebody asks a question on Stack Overflow, they receive a recommendation to attach at least three tags to the question. The body part of the thread contains the description
of the question. Whenever a question is answered, the answer appears in the answers section of the thread. Other developers can also ask further clarifying questions or comment on the question or answers posted up to that point. 

\vspace{0.2cm}\noindent{\it Twitter and Software Engineering}:
Twitter is currently one of the most popular microblogging sites in the world. On Twitter, a user can post short messages (a.k.a. {\em tweets}) broadcasted to all other Twitter users who are following the user. Twitter allows a user to {\em follow} another user, which means the latter subscribes to all the tweets of the user he/she is following. Users also have an option of reposting the tweets posted by others -- an activity known as {\em retweeting}. Twitter also allows users to mark favorite tweets, which conveys their interest in the content of a tweet.

By virtue of its simple design and easy-to-use functionality, Twitter has become a powerful medium for information sharing and dissemination. It started as a social networking medium but has nowadays become one of the important sources of information for people to keep up-to-date with the latest news and information about their domains of interest, to share and promote knowledge, and to keep in touch with their family and friends~\cite{KwakLPM10}. Twitter influences many communities including the software engineering community as highlighted by many prior studies~\cite{singer2014software,Bougie11,Wang13,Tian12does}. Various techniques have been proposed recently to mine software engineering relevant information from Twitter~\cite{sharma2015nirmal,sharma2017harnessing,williams2017mining,guzman2017little}.

\vspace{0.2cm}\noindent{\it YouTube and Software Engineering}: YouTube is a website where anybody can share videos \cite{chenail2008youtube}. It has over 1 billion users and generates billions of views daily~\cite{aboutyoutube}. YouTube has also evolved into a knowledge sharing resource, where people can share informational videos, follow other users and comment on videos. Thus it provides people with resources to share information, learn new knowledge, as well as get and provide feedback.

Software developers also use YouTube for sharing information as well as learning~\cite{macleod2015code,ponzanelli2016too}. MacLeod et al. found that developers share videos detailing information they wished they had found earlier~\cite{macleod2015code}. The videos mainly relate to sharing knowledge about development experiences, implementation approaches, design pattern application, etc. Other work focuses on extracting relevant information for developers from YouTube, which is a challenging task given the large size of videos. Tools to help developers find relevant content from software engineering videos have been proposed in~\cite{ponzanelli2016codetube,yadid2016extracting}. Poch\'e et al. analyzed user comments related to software engineering videos posted on YouTube~\cite{poche2017analyzing} and proposed a technique for finding relevant comments.

\subsection{Transfer Representation Learning and Word Embeddings}

Representation learning can be described as learning representations of data that make it easier to extract useful information when building classifiers or other predictors~\cite{bengio2013representation}. In the field of Natural Language Processing (NLP) applications, distributed word representations are one of the applications of representation learning. Distributed word representations, i.e., word embeddings, have been widely applied in various text mining and natural language processing tasks. 

Word embeddings represent words in a low dimensional continuous space, to convey semantic and syntactic information~\cite{li2015word}. One of the most popular word embedding techniques is Word2Vec, which uses a shallow neural network to reconstruct contexts of words. Mikolov et al.~\cite{mikolov2013efficient,mikolov2013distributed} proposed two word embedding models Continuous Bag-of-Word (CBOW) and the skip-gram model which have been widely adopted due to their effectiveness and efficiency. For CBOW, a neural network is trained to predict a word based on its surrounding words. In this architecture, the continuous value vector for a word is the vector that is input to the last layer in the network after we input its surrounding words to the network. For skip-gram, a neural network is trained to predict surrounding words based on the current word. In this architecture, the continuous value vector for a word is the vector that is output by the first layer in the network. It has been shown that the embedding vectors produced by these models preserve the syntactic and semantic relations between words under simple linear operation. For example, the resultant vector of the following arithmetic operation (vector of brother - vector of man + vector of woman) is similar to the vector of sister. This is related to analogical reasoning where brother is to sister as man is to woman, which is encoded in the vector representation learned by Word2Vec.

In machine learning, many methods perform well under the common assumption that the training and test data are drawn from the same feature space and the same distribution. In many contexts, this assumption may not hold. For example, we attempt to solve a classification problem in a domain that does not have enough training data, but we have sufficient data in other related domains. In this case, knowledge transfer or transfer learning would be useful to solve the classification problem~\cite{pan2010survey}. In the context of representation learning, transfer representation-learning is where rich representations are learnt in a source platform with the aim of transferring them to different target platforms~\cite{andrews2016transfer}.

\section{Representation Learning from Sofware-Development-Specific Platforms}
\label{sec.approach} 

Our work is related to transfer representation-learning, where rich representations are learnt from a software-development-specific platform, and leveraged in a different target platform. To represent knowledge in the source platform, we build a word embedding model that represents each word as a low-dimensional vector such that words that are similar in meaning are associated with similar vectors. Word embedding models have successfully been applied in various natural language processing (NLP) tasks, such as in~\cite{ye2016word,xu2016domain,chen2016learning}. 

A recent finding by Mou et al.~\cite{mou2016transferable} shows that the transferability of neural NLP models depends largely on the semantic relatedness of the source and target tasks. Therefore, since our target tasks are related to the extraction of knowledge relevant to software development, we need to define a source platform that contains rich software engineering content. In this work, we choose two source platforms (Software Engineering Stack Exchange and Stack Overflow), and compare the performance of models built from the two platforms.

Our approach consists of two stages: representation learning from a source platform, and model building for a target platform. In the first stage, we learn a word embedding representation from a software-development-specific platform. In the second stage, we leverage the word embedding model built from the platform to resolve tasks in the target platform. Figure \ref{figFramework} shows the overall architecture of our proposed framework. 

\begin{figure}[!htb]
	\centering
	\scalebox{0.40}{\includegraphics{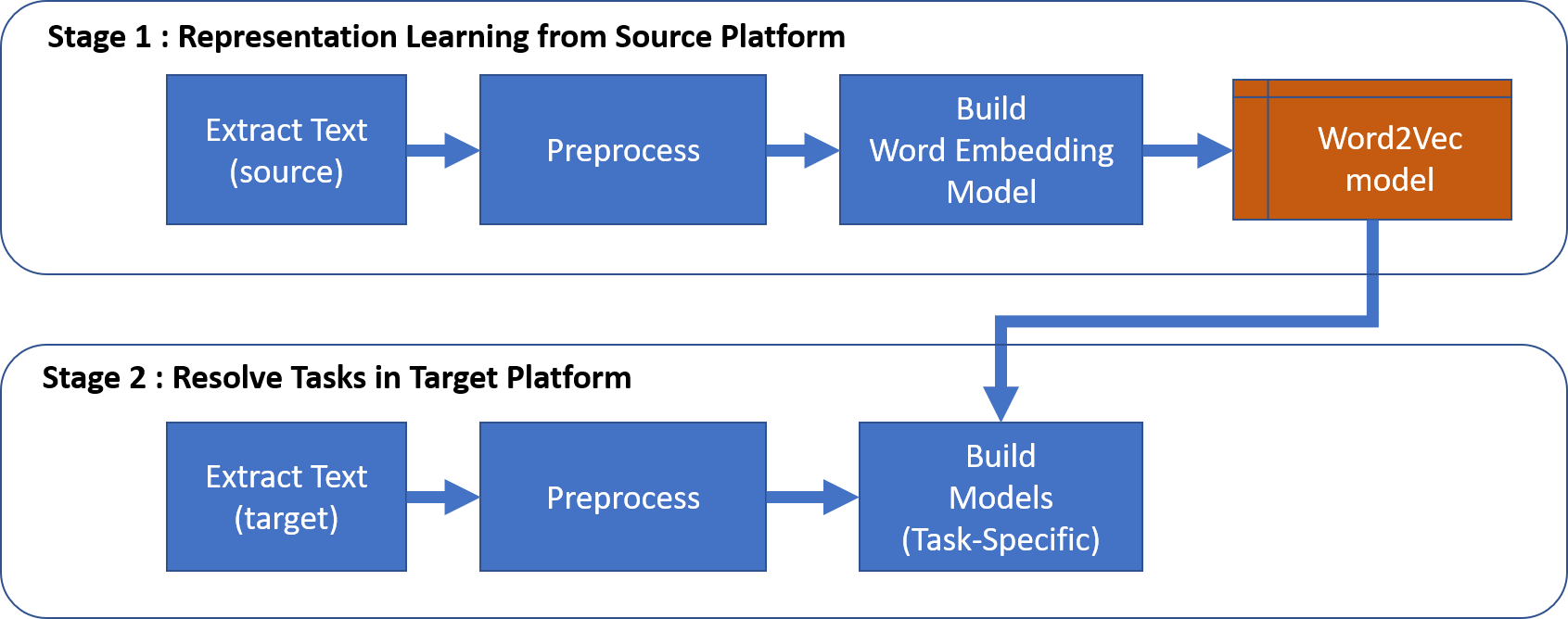}}
	\caption{Overall approach}
	\label{figFramework}       
\end{figure}

We describe each stage of our proposed approach as follows:

\vspace{0.2cm}\noindent{\bf Stage 1: Representation Learning from Source Platform}

While most research done on Q\&A sites is based on Stack Overflow data (e.g.~\cite{sharma2015nirmal,zhangrecommending}), we believe that Software Engineering Stack Exchange (StackExchange-SE) is also a good source for software engineering related terms. Therefore, we use text data extracted from the two sites, and build two different models: \texttt{SIEVE\_SO} which is based on \textit{Stack Overflow} and \texttt{SIEVE\_SE} which is based on \textit{Stack Exchange} data. 

The StackExchange-SE dataset is publicly available on the Stack Exchange data dump site.\footnote{http://archive.org/download/stackexchange\label{stackexchangedumpurl}} We use the following two files: Posts.7z and Comments.7z. Posts.7z contains the title and body of posts (i.e., questions and answers) that appear on Stack Exchange. Comments.7z contains comments that users give to the questions and answers on Stack Exchange. Our Stack Exchange dataset contains a total of 149,478 posts, 409,740 comments, and 46,246 titles generated in a time period spanning from September 2010 to August 2017. We combined all of the posts, comments and titles for learning word embeddings from this dataset.


We used the Stack Overflow data dump provided by previous  work by Sharma et al.~\cite{sharma2015nirmal}. The data was also taken from the data dump site.\textsuperscript{\ref{stackexchangedumpurl}} They extracted the questions and answers from the Posts.7z file, and user's comments from Comments.7z. These files contain content posted on Stack Overflow from September 2008 to September 2014. There are a total of 7,990,787 titles, 21,736,594 posts (questions and answers), and 32,506,636 comments. Since there are too many posts and comments to efficiently process the data, to reduce the time it takes to learn a model, they randomly selected 8 million posts and comments from the data dump. We use this randomly selected data and combine all of the posts, comments and titles.


Before we build the word embeddings model, we performed the following text preprocessing for both datasets:
\begin{enumerate}
	\item Parse the posts into sentences, since we want to train word embeddings at sentence-level. We use NLTK\textsc{\char13}s \textit{punkt} tokenizer\footnote{http://www.nltk.org} for sentence splitting.
	\item Remove all HTML tags since they do not contain useful information for word embeddings.
	\item Remove all special characters (e.g., symbols, punctuations, etc.) and words that contain only numbers.
	\item Change all words to their lower case.
\end{enumerate}


We chose the continuous skip-gram Word2Vec model proposed by Mikolov et al.~\cite{mikolov2013efficient} We use the Word2Vec implementation in Gensim\footnote{https://pypi.org/project/gensim/}. We set the parameters according to Mikolov et al.~\cite{mikolov2013efficient}: context windows size to 5, dimension to 300, batch size to 50, negative sampling to 10, minimum word frequency to 5 and iterations to 5. The output of the model is a dictionary of words, each of which is associated with a vector representation. Table~\ref{sec.tabstat} includes statistics on the generated word embeddings learned from the datasets.

\begin{table}[!htb]
	\caption{ Statistics of datasets and word embeddings extracted from Stack Overflow (\texttt{SIEVE\_SO}) and StackExchange-SE (\texttt{SIEVE\_SE}) }
	\centering	
	\scalebox{1}{
		\begin{tabular}{|l|r|r|}
			\hline 
			\multicolumn{1}{|c|}{\textbf{ }} & \multicolumn{1}{|c|}{\textbf{StackExchange-SE}} &  \multicolumn{1}{c|}{\textbf{Stack Overflow}} \\
			\hline
			\hline Number of Posts+Comments & 605,464 & 8,000,000  \\
			\hline Number of Sentences (after preprocessing) & 1,884,959 & 5,007,411
			\\
			\hline Size of Vocabulary in word vector & 232,953 & 275,103
			\\
			\hline					
		\end{tabular}
	}
	\label{sec.tabstat}
\end{table}

\vspace{0.2cm}\noindent{\bf Stage 2: Model Building for Target Platform}

Our goal is to leverage knowledge extracted from software-development-specific platforms and apply it to open-domain platforms. In order to examine the learned word embeddings representation in stage 1, we utilize the word embeddings in two different use cases. In the first use case, we aim to resolve the task of finding tweets related to software engineering. In the second use case, we leverage the word embeddings to classify user comments on YouTube coding tutorial videos. We discuss each of the use cases further in the next sections.

\section{Finding Relevant Tweets Using Word Embeddings}
\label{sec.task1} 
 
In this section, we show how our approach can be used for the task of finding tweets related to software engineering. Researchers have found that developers use Twitter to support their professional activities by sharing and discovering various information from microblogs, e.g., new features of a library, new methodologies to develop a software system, opinions about a new technology or tools, etc.~\cite{sharma2015nirmal} However, due to various topics posted on Twitter, it becomes a challenge to find interesting software-related information on Twitter. To overcome this problem, Sharma et al.~\cite{sharma2015nirmal} proposed a language-model based approach and used the model to rank tweets based on their relevance to software engineering. We will use the proposed model as a baseline, along with other baselines. We aim to answer the following research question:

\vspace{0.2cm}\noindent{\bf RQ1. How effective is our approach at the task of finding software related tweets?}


\subsection{Approach}
Figure~\ref{fig.task1_framework} shows an instance of our proposed approach for the task of finding software development-related tweets, by utilizing word embeddings trained from a source platform. In general, we formulate the task of finding software-related tweets as a ranking problem, i.e., ranking the tweets in the order of their similarity scores with selected sentences from the source platforms. We follow these steps:

\begin{figure}[!htb]
	\centering
	\scalebox{0.40}{\includegraphics{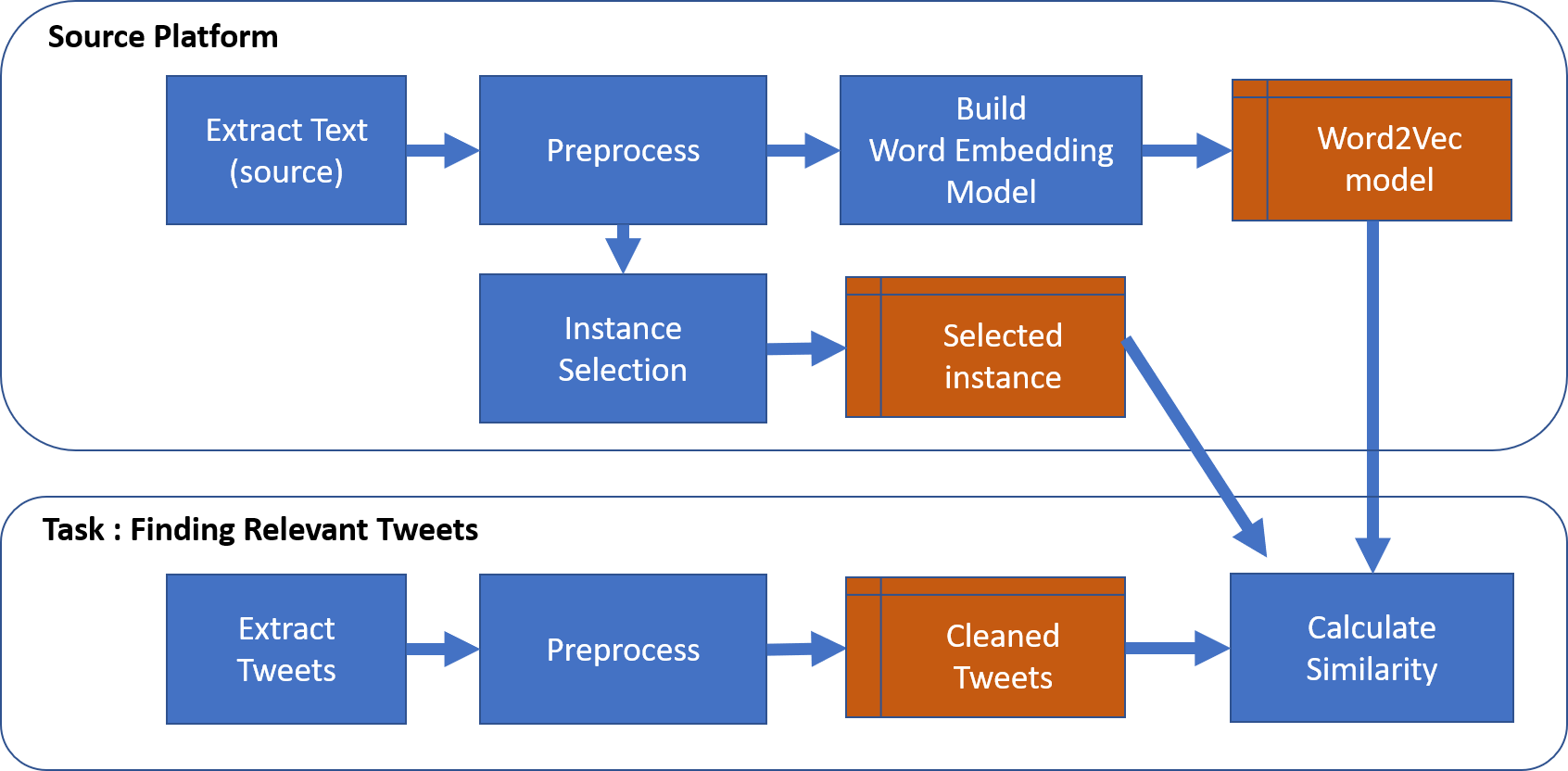}}
	\caption{Our approach for finding software-related tweets}
	\label{fig.task1_framework}       
\end{figure}

\vspace{0.2cm}\noindent{\bf Step 1: Instance Selection}

In our approach, selecting instances (i.e., sentences) from the source platform is an important task, since we will use these selected sentences to rank the tweets based on a similarity measure. Sentences extracted from the source platform (StackExchange-SE/Stack Overflow) are considered as software-related. However, some of the sentences may have different characteristics with Twitter. Therefore, we use the following heuristic methods to select suitable sentences from a source platform: 
\begin{enumerate}
	\item We select sentences that have a length of no more than 140 characters which corresponds to a tweet's maximum length.
	\item Among these selected sentences, we randomly sample  sentences. By default, we sample 1000 sentences. We believe that this sampled set should be enough to represent sentences that contain software-related terms.
\end{enumerate}

\vspace{0.2cm}\noindent{\bf Step 2: Preprocess Tweets}

We use the Twitter dataset provided by Sharma et al.~\cite{sharma2015nirmal}. The tweets are preprocessed by removing punctuation marks and URLs, and all words are changed into lowercase.  

\vspace{0.2cm}\noindent{\bf Step 3: Calculate Similarity}

To measure similarity between tweets and selected sentences taken from the source platforms, we need to implement the same representation for both texts. Because sentences (or tweets) have different lengths, we need to use a fixed-length vector to represent them. To build the vector representation, we leverage the word embeddings learned from a source platform. The model consists of word vectors that have a dimension of 300 as mentioned in Section~\ref{sec.approach}. We follow these steps:

\begin{enumerate}
	\item For each sentence or tweet in the dataset, we tokenize it into words. \item For each word, we look up its weight from the word embeddings model. If a word does not exist in the model, we can either ignore that word, use a vector whose values are all 0 to represent it, or use the average of the embeddings from words having the lowest frequency in the model. By default, we ignore the word that not exist in the model. The result is a 300-dimension vector of real values taken from the word embeddings model.
	\item We represent the sentence into a fixed-length vector. There are different ways to obtain text representation from word embeddings. The most common methods use the maximum, minimum, or average of the embeddings of all words (or just the important words) in a sentence~\cite{socher2013recursive}. In this case, we take the average of the word embeddings of all words within the text, following~\cite{kenter2015short}. At the end, we have a word vector of real values with dimension of 300 for each tweet or sentence.
\end{enumerate}

For each tweet, we calculate similarity between the vector representations of tweets and the vector representations of each of selected sentences in the source platform. We then rank the tweets based on their similarity scores. The higher the scores, the more likely the tweet contains software-related contents. To calculate similarity between two word vectors, we use cosine similarity. Cosine similarity is a measure of similarity between two vectors (in this case, vector of text representation) that measures the cosine of the angle between them. Given a tweet $T$ and a selected sentence $S$, that are represented by two word vectors $wv_{tweet}$ and $wv_{so}$, we define their semantic similarity as the cosine similarity between their word vectors:
\[
similarity(T,S) =  \frac{wv_{tweet}^T . wv_{so}}{||wv_{tweet}|| || wv_{so}||}
\]

\subsection{Dataset and Baselines}
\vspace{0.2cm}\noindent{\bf Dataset.}
For the Twitter dataset, we use the same dataset used by Sharma et al.~\cite{sharma2015nirmal}. The dataset consists of around 6.2 million tweets downloaded through the Twitter REST API. To collect tweets, they first obtained a set of microbloggers that are likely to generate software-related contents. They started with a collection of 100 seed microbloggers who are well-known software developers. Next, they analyzed the follow links of these microbloggers to identify other Twitter accounts that follow or are followed by at least 5 seed microbloggers. After they had identified the target microbloggers, they downloaded tweets that were generated by these individuals. They then performed preprocessing on the collected tweets such as removing punctuation marks and URLs, and changed all words into lowercase.

\vspace{0.2cm}\noindent{\bf Baselines.}
We used several baselines to show the effectiveness of our approach. First, we compared our proposed approach against NIRMAL~\cite{sharma2015nirmal}, since we used the same dataset as their work. Next, since our models are trained on software-development-specific platforms, we compared the models with a within-platform model trained from the target platform (Twitter). To show the effectiveness of the Word2Vec-based models that we use, we compared the models with a model that uses Term Frequency $-$ Inverse Document Frequency (td-idf) vectors generated from a source plarform. Tf-idf technique has been widely used in other software-engineering-related information retrieval tasks such as in~\cite{de2014labeling,palomba2016textual}. We briefly describe the baselines as follows:
\begin{enumerate}
	\item \textbf{NIRMAL by Sharma et al.} We used this approach as the main baseline, since the approach is the state-of-the-art in the task of ranking software-related tweets. This approach builds an N-gram language model by using SRILM~\cite{stolcke2002srilm}, a language modeling toolkit. NIRMAL learns a language model from Stack Overflow data. NIRMAL then uses the learned model to compute the perplexity score of each tweet. The lower the perplexity score, the more likely the tweet is software related. NIRMAL then ranks the tweets in ascending order of their perplexity scores and returns a ranked list.	
		
	\item \textbf{Term Frequency $-$ Inverse Document Frequency (\textit{tf-idf}).}
	In this approach, instead of using vectors generated by word embeddings, we used td-idf vectors generated from a source plarform. We built two variants of tf-idf vectors: one from Stack Overflow  posts, and one from Stack Exchange posts. Term frequency (tf) is the number of times a word occurs in a given sentence, accompanied with a measure of the term scarcity across all the sentences, known as inverse document frequency (idf). Before constructing the vectors, we performed stemming using Porter Stemmer~\cite{porter1980algorithm} and removed English stopwords. To remove stopwords, we used stopwords listed in the Python NLTK library\footnote{https://www.nltk.org/}.
	 
	\item \textbf{Word2Vec trained on Twitter.} 
	We consider this approach as a {\em within-platform} baseline, since we leverage knowledge extracted from Twitter itself as the target platform. We trained a skip-gram word embeddings model using the set of parameters advised by Mikolov et al.~\cite{mikolov2013efficient}: context windows size of 5, dimensions of 300, batch size of 50, negative sampling to 10, minimum word frequency of 5 and iterations of 5 - the same set of parameters that are used by our proposed approach.
	
\end{enumerate}

\subsection{Experiments and Results}

\vspace{0.2cm}\noindent{\bf Experiments Setting.}
We conducted experiments to answer RQ1 and evaluated the effectiveness of our approach as compared to the baselines. After following the steps in our proposed approach, we ranked the tweets based on similarity scores between the tweets and selected instances taken from a source platform. We investigated three different word embeddings models trained from Stack Overflow, StackExchange-SE and Twitter, and one non-word embeddings model (tf-idf).

As an evaluation metric, we used accuracy@K, which is defined as the proportion of tweets in the top-K positions that are software-related. We manually evaluated the top-K tweets ranked by their similarity scores. We asked two labelers who have master's degrees in Computer Science to manually label the tweets, either as \textit{"relevant"} or \textit{"not relevant"} to software engineering. For our final ground truth, we labeled a particular tweet as "relevant" only if both labelers agreed that the tweet is software-development-related. We used Cohen's Kappa to measure inter-rater reliability for the labeling task. We obtained 
a Kappa value of 0.78 for labeling \texttt{SIEVE\_SE} and a Kappa value of 0.68 for labeling \texttt{SIEVE\_SO}  -- following Landis and Koch's interpretation~\cite{landis1977measurement}, this value indicates substantial agreement.

\vspace{0.2cm}\noindent{\bf Results.}
The results of our experiments are shown in Table~\ref{sec.rq1.tab1} and Figure~\ref{fig.task1_result}. Overall, the word embeddings model trained on Stack Exchange performed best, except for accuracy@10, where the tf-idf based approach performed best. Word embeddings models trained on platforms that contain rich software-development-related knowledge (Stack Exchange and Stack Overflow) performed better as compared to the baselines.

\begin{table}[hbt]
	\caption{ Accuracy@K results of different approaches in our experiments (best results are in bold).}
	\centering	
	\scalebox{1}{
		\begin{tabular}{|l|c|c|c|c|c|c|c|c|c|c|}
			\hline \textbf{Approach } & \textbf{acc@10} & \textbf{acc@50} & \textbf{acc@100} & \textbf{acc@150}  & \textbf{acc@200} \\
			\hline
			\hline \texttt{Nirmal} & 0.900 & 0.820 & 0.720 & 0.707 & 0.695 \\
			\hline \texttt{TF-IDF (Stack Overflow)}& \textbf{1.000} & 0.540 & 0.730 & 0.693 & 0.575 \\
			\hline \texttt{TF-IDF (Stack Exchange)}& \textbf{1.000} & 0.560 & 0.400 & 0.347 & 0.310 \\
			\hline \texttt{Word2Vec (Twitter)}& 0.400 & 0.220 & 0.260 & 0.260 & 0.235 \\
			\hline \texttt{SIEVE\_SO}& 0.900 & 0.880 & 0.870 & 0.847 & 0.800 \\
			\hline \texttt{SIEVE\_SE}& 0.900 & \textbf{0.980} & \textbf{0.970} & \textbf{0.940} & \textbf{0.925} \\
			\hline		
		\end{tabular}
	}
	\label{sec.rq1.tab1}
\end{table}

\begin{figure}[!htb]
	\centering
	\scalebox{0.95}{\includegraphics{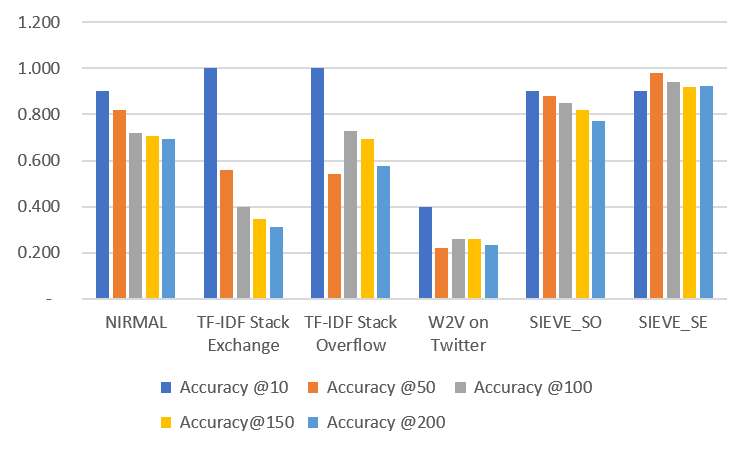}}
	\caption{Comparison of Accuracy@K achieved by different approaches}
	\label{fig.task1_result}       
\end{figure}

Based on our experiments, the performance of Word2Vec trained on Twitter was lower than the other Word2Vec models. The low scores achieved by the Word2Vec model trained on Twitter data can be attributed to the content of tweets that are mostly not related to software development. 

While the tf-idf-based approach performed best when ranking the top-10 most relevant tweets, the  performance degrades significantly when ranking top 50, and fluctuates when ranking the top 100, 150 and 200 tweets. Figure~\ref{fig.task1_tweets} shows a box-plot diagram, describing the word count of the top 200 tweets returned by various approaches. We found that, in the top-200 tweets ranked by the tf-idf approach, the tweets mostly contain word stem "use", such as "\textit{What is the use of c}", "\textit{Java MySQL Insert Record using Jquery}", "\textit{@shayman I used to work there"}. On the other hand, the top 200 tweets returned by SIEVE\_SE contain more diverse vocabulary and tend to be lengthy, such as "\textit{I still very much admire all the work put into TinyMCE Building a RTE is one of the most gruesome things you can do in a browser}," "\textit{@Youdaman yes angular is very minimal in the amount of code and glue you need to do specially if you use a RESTful service}". This finding highlights the benefit of leveraging word embedding models to learn feature representation from a rich software-related platform.

\begin{figure}[!htb]
	\centering
	\scalebox{0.95}{\includegraphics{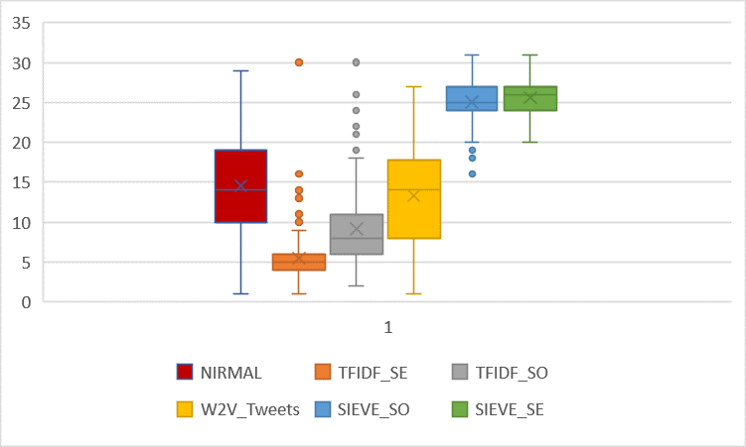}}
	\caption{A box plot diagram representing word count of tweets returned by various approaches}
	\label{fig.task1_tweets}       
\end{figure}

\begin{tcolorbox}
	
	RQ1: Two variants of SIEVE (\texttt{SIEVE\_SE} and \texttt{SIEVE\_SO}) are able to find software-related tweets with accuracy of 0.800 - 0.980.
	
\end{tcolorbox}

\section{Finding Informative Comments on YouTube Using Word Embeddings}
\label{sec.task2} 

The objective of this task is to analyze user comments for YouTube coding tutorial videos. Important users' questions and concerns can then be automatically classified in order to help content creators to better understand the needs and concerns of their viewers, as described in work by Poch\'e et al.~\cite{poche2017analyzing} They categorized the comments into two general categories: informative vs. non-informative (which corresponds to other miscellaneous comments). We aim to answer the following research question:

\vspace{0.2cm}\noindent{\bf RQ2. How effective is our approach at the task of finding informative comments on YouTube?}

\subsection{Approach}
Figure~\ref{fig.task2_framework} shows our proposed approach for the task of finding informative comments on YouTube, by utilizing word embeddings models trained on the source platforms (StackExchange-SE and Stack Overflow).  We formulate this task as a binary classification problem, where a comment can be either informative or non-informative with regards to the video content. In order to build a classifier for this task, we need to represent the YouTube comments into a feature representation. We leverage the word embeddings learned from a source platform. The model consists of word vectors that have 300 dimensions as mentioned in Section~\ref{sec.approach}. We build vectors to represent the comments, by following these steps:
\begin{figure}[!htb]
	\centering
	\scalebox{0.40}{\includegraphics{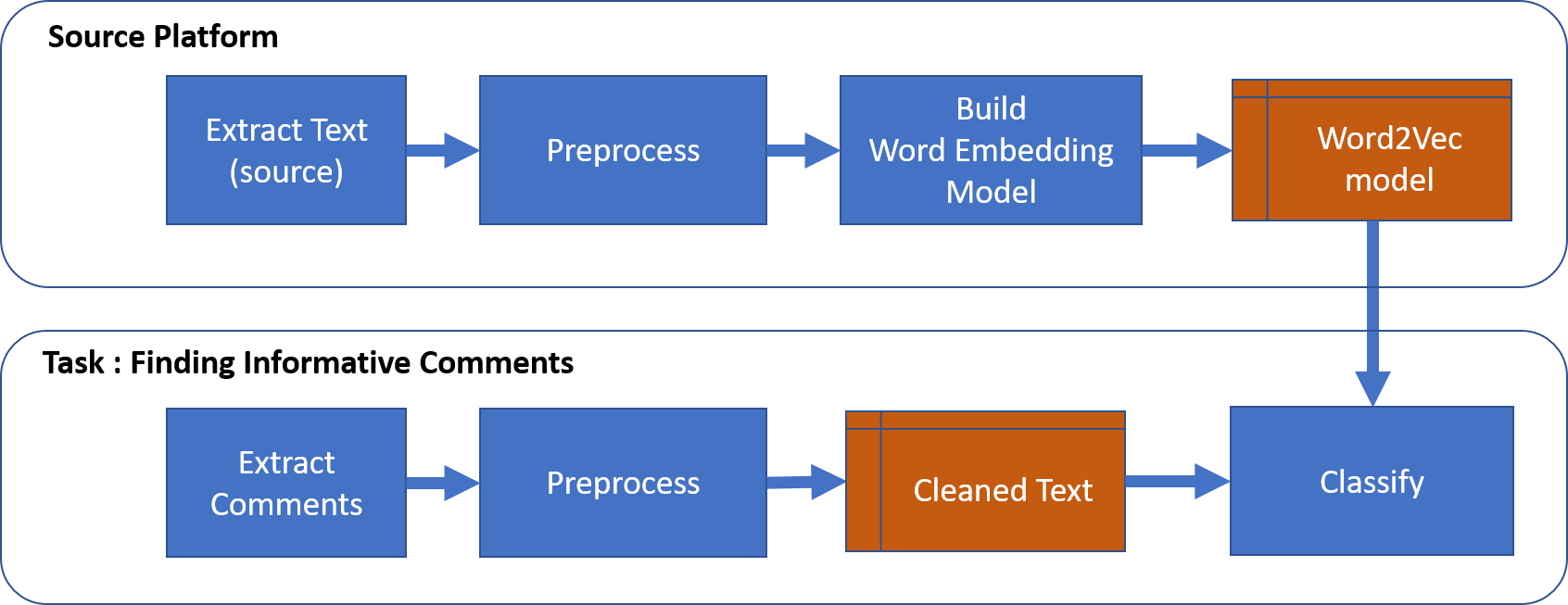}}
	\caption{Approach for finding relevant comments on YouTube}
	\label{fig.task2_framework}       
\end{figure}

\begin{enumerate}
	\item For each comment, we tokenize it into words, remove words that contain only numbers, and change all words into lowercase. 
	\item Next, for each word in the comment, we look up its vector value taken from the word embeddings model. We ignore a word if it does not exist in the word embeddings model. We take the average of the word embeddings of all words within the text, following~\cite{kenter2015short}. At the end, we have a word vector of real values with dimension of 300 for each comment.	
\end{enumerate}

\subsection{Dataset and Baselines}

\vspace{0.2cm}\noindent{\bf Dataset.}
We used the dataset provided by Poch\'e et al.\footnote{http://seel.cse.lsu.edu/data/icpc17.zip} The dataset consists of 6,000 YouTube comments sampled from 12 different coding tutorial videos. The data was collected on Sep 6, 2016. They collected a total of 41,773 comments from all videos. They used YouTube Data API3\footnote{https://developers.google.com/youtube/v3/} to retrieve the comments. This API extracts comments and their metadata, including the author\textsc{\char13}s name, the number of likes, and the number of replies. Finally, 500 comments were sampled from the videos. Based on a manual classification process, around 30\% of the comments were found to be informative, meaning that the majority of comments are basically not related to the content.

\vspace{0.2cm}\noindent{\bf Baselines.}
Since we used the same dataset and experiment setting as Poch\'e et al.'s work, we used their approach as the first baseline. To show the effectiveness of our models that are trained on software-development-specific platforms, we compared the models with a within-platform model trained from YouTube comments, and another cross-platform pretrained model that was learned from more general contents. We briefly describe the baselines as follows:
\begin{enumerate}
	\item \textbf{Normalized Term-Frequency (as proposed by Poch\'e et al.~\cite{poche2017analyzing}).} 
	In order to automatically identify content-relevant comments, Poch\'e et al. investigate the performance of two classification algorithms: Naive Bayes (NB) and Support Vector Machines (SVM). They performed text preprocessing on the dataset, by stemming and removing stopwords. They also remove words that appear in one comment only since they are highly unlikely to carry any generalizable information. As feature representation, they use normalized term frequency (tf) of words in their documents. They found that their SVM classifier performs better than Naive Bayes. They also experimented with different combinations of data preprocessing such as stemming and removing stop-words, and found that the best result was achieved without stemming and stop-word removal. 
	
	\item \textbf{Word2Vec trained from YouTube Comments.}  
	We consider this baseline as a {\em within-platform} baseline, since we leverage knowledge extracted from the target platform itself (i.e., YouTube comments). We built a skip-gram word embeddings model from this dataset with the same set of parameters used by our proposed approach.
	
	\item \textbf{Pretrained Word2Vec on GoogleNews.} 
	We used a pretrained word embedding model on GoogleNews\footnote{https://code.google.com/archive/p/word2vec/} which is an alternative cross-platform pretrained model as another baseline. The model contains 300-dimensional vectors for 3 million words and phrases, which was trained on part of Google News dataset (about 100 billion words).

\end{enumerate}

\subsection{Experiments and Results}

\vspace{0.2cm}\noindent{\bf Experiments Settings.}
We conducted experiments to answer RQ2 and evaluated the effectiveness of our approach as compared to the baselines. We used Support Vector Machines (SVM) as the classification algorithm, since this algorithm performs better in Poch\'e et al.'s work~\cite{poche2017analyzing}. To enable a fair comparison, we used the same implementation of SVM (inside Weka\footnote{https://www.cs.waikato.ac.nz/ml/weka/}) for classification. For the kernel function, in Poch\'e et al's work, the best results were obtained using the universal kernel. Therefore, we also used the universal kernel in our experiment. To validate the result, we used 10-fold cross validation. With this technique, the dataset was first partitioned randomly into 10 partitions of equal size. Afterwards, one of the partitions was selected as validation set while the remaining partitions are used for training. The process was repeated 10 times with a different partition being selected as validation set, ensuring that the entire dataset was used for both training and validation, and each entry in the dataset was used for validation exactly once.

To measure the effectiveness of our approach, we used the same metrics as Poch\'e et al.'s study (i.e., Precision, Recall and F-measure). F-measure is the harmonic mean of precision and recall, and it is used as a summary measure to evaluate if an increase in precision (recall) outweighs a reduction in recall (precision). These metrics are calculated based on four possible outcomes of each comment in an evaluation set: True Positive (TP), True Negative (TN), False Positive (FP) and False Negative (FN). TP corresponds to the case when a comment is correctly classified as an informative comment; FP corresponds to the case when a non-informative comment is wrongly classified as an informative comment; FN is when a comment is wrongly classified as a non-informative comment; TN is when a non-informative comment is correctly classfied as such. The formulas to compute precision, recall, and F-measure are shown below: 

\[ Precision=\frac{\#TP}{\#TP+\#FP} \]
\[ Recall =\frac{\#TP}{\#TP+\#FN} \]
\[ FMeasure =2 \times \frac{Precision \times Recall}{Precision+Recall} \]

\vspace{0.2cm}\noindent{\bf Results.}
The results of our approach as compared to the baselines are shown in Table~\ref{sec.rq2.tab2} and Figure~\ref{fig.task2_result}. The results showed that the best performance (in terms of precision, recall, and F-measure) was achieved by using word embeddings model trained on StackExchange-SE data.  

\begin{table}[hbt]
	\caption{ Performance of different approaches for classifying informative comments. }
	\centering	
	\scalebox{1.1}{
		\begin{tabular}{|l|c|c|c|}
			\hline \textbf{Approach } & \textbf{Precision} & \textbf{Recall} & \textbf{F-Measure} \\
			\hline
			\hline \texttt{NTF (Poch\'e et al.)} & 0.790 & 0.750 & 0.770  \\
			\hline \texttt{Word2Vec (GoogleNews)} & 0.859 & 0.861 & 0.859  \\
			\hline \texttt{Word2Vec (YouTube Comments)} & 0.829 & 0.833 & 0.830  \\
			\hline \texttt{SIEVE\_SO} & 0.868 & 0.870 & 0.869 \\
			\hline \texttt{SIEVE\_SE} & \textbf{0.872} & \textbf{0.874} & \textbf{0.873} \\
			\hline					
		\end{tabular}
	}
	\label{sec.rq2.tab2}
\end{table}

The results also showed that by using word embeddings as feature representation, the performance of the classifiers can be improved by up to 10.3\% in terms of F-measure, as compared to the normalized-tf based approach proposed by Poch\'e et al. 
Among the four word embeddings models used in our experiment, models trained on Stack Exchange and Stack Overflow performed best. This finding justifies the importance of choosing a source platform that is more relevant to a target task. Even though the corpus' size is less as compared to GoogleNews data, Stack Exchange and Stack Overflow data contains more software-development-specific contents than GoogleNews, and this explains the improved performance. 

\begin{figure}[!htb]
	\centering
	\scalebox{0.9}{\includegraphics{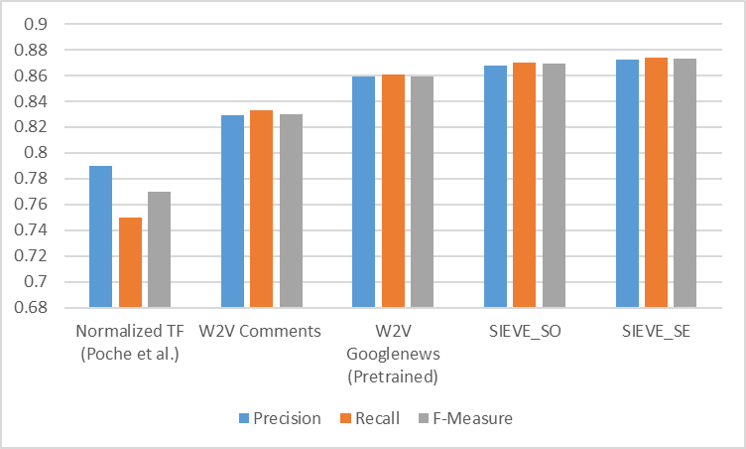}}
	\caption{Comparison of Precision, Recall and F-measure achieved by different approaches}
	\label{fig.task2_result}       
\end{figure}

\begin{tcolorbox}
	
	RQ2: Two variants of SIEVE (\texttt{SIEVE\_SE} and \texttt{SIEVE\_SO}) performed better than various baselines with F-measure score of 0.874 and 0.869 respectively
	
\end{tcolorbox}

\section{Threats to Validity}
\label{sec.threats}
We present the potential threats to the validity of our findings. The threats include threats to internal, external, and construct validity.

\vspace{0.2cm}\noindent{\bf Threats to internal validity.} These threats are related to potential errors that may have occurred when performing the experiments and labelling. Internal threats might stem from the tools we used in our analysis. We used Gensim\footnote{https://pypi.org/project/gensim/}, a popular Python module for machine learning to build word embeddings that has also been used in many previous studies related to word embeddings. For machine
learning and classification tools, we used Weka\footnote{https://www.cs.waikato.ac.nz/ml/weka} which has been extensively used in the literature and has been shown to generate robust results for various applications. Potential errors might also occur when labelling our dataset. To label tweets as software related or not, we asked two labelers with experience in programming, and with degrees in Computer Science. We believe the labelers have enough expertise to judge if a tweet is software-related or not. 
 
\vspace{0.2cm}\noindent{\bf Threats to external validity.} These threats refers to the generalizability of our results. To mitigate these threats, we have considered two source domains (Software Engineering Stack Exchange and Stack Overflow), two target domains (Twitter and YouTube), two tasks (relevant tweet identification and informative comment classification), and two  settings (ranking and classification). 

\vspace{0.2cm}\noindent{\bf Threats to construct validity.} These threats are related to the suitability of the evaluation metrics that we use for analyzing the result. We use the same evaluation metrics used to evaluate previous studies~\cite{sharma2015nirmal,poche2017analyzing} to enable fair comparisons (i.e., Accuracy@k, Precision, Recall and F-measure).  Therefore, we believe that the threat to construct validity is minimal. 
\section{Related Work}
\label{sec.related}
In this section, we describe work related to our study.
\subsection{Developer Information Channels} 
 In the past few years, there has been a substantial amount of  work which has analyzed tools or channels used by software developers. Storey et al. found that software developers use many communication tools and channels in their software development work~\cite{storey2014r,storey2017social}. They found that a lot of knowledge is embedded in these tools and channels which also encourages a participatory culture in software development. Their work also highlights the challenges faced by developers while using  these channels. In the paragraphs below, we discuss work related to channels such as Software Engineering Stack Exchange and Stack Overflow (software question-and-answer sites), Twitter, and Youtube, as these are the domains we have considered in this work.

Stack Overflow has received much attention in recent years in the software engineering research community. Barua et al. analyzed in detail the topics and trends among discussions on Stack Overflow by applying Latent Dirichlet Allocation (LDA)~\cite{barua2014developers}. They found that  the topics which interest developers range from jobs to version control systems to C\# syntax etc. Their analysis showed that web development, mobile applications, Git, and MySQL were the topics  that were gaining  the most popularity over time.  Vasilescu et al. explore how the developer's use of Stack Overflow and GitHub relates to each other~\cite{vasilescu2013stackoverflow}. Many empirical studies have focused on understanding  and modeling questions and/or answers on Stack Overflow. Asaduzzaman et al. found  that some questions go unanswered on Stack Overflow as the questions may be short, unclear, too hard, etc.~\cite{asaduzzaman2013answering}.  Rahman et al. developed a prediction model based on behavior, topics, and popularity
 of a question to determine unresolved questions~\cite{rahman2015insight}. Ponzanelli et al. have performed studies to analyze the quality of questions on Stack Overflow~\cite{ponzanelli14understanding} and also proposed an approach to detect low-quality questions~\cite{ponzanelli2014improving}. 
 Treude et al. did an analysis of how programmers ask and answer questions on Stack Overflow~\cite{treude2011programmers}. How the crowd generates valuable documentation on Stack Overflow  and ways of measuring it have been discussed in~\cite{parnin2012crowd,parnin2011measuring}.   A lot of tools have also been proposed which can help developers in their usage of Stack Overflow. Many techniques have been proposed for solving the problem of tag prediction for questions on Stack Overflow~\cite{wang2014entagrec,zhou2017scalable,cai2016greta}. 
 Seahawk, an Eclipse plugin was proposed to integrate Stack Overflow knowledge within the IDE~\cite{ponzanelli2013seahawk,bacchelli2012harnessing}. Identification of opinionated sentences from Stack Overflow data and subsequent aspect identification have been proposed recently by Uddin et al.~\cite{uddin2017automatic,uddin2017opiner}.

\textit{Stack Exchange} was first explored from a software engineering perspective by Begel et al.~\cite{begel2013social}. This work explored what kinds of service were provided by Stack Exchange, what challenges they face, and how people benefit from the service. Possnet et al. did an empirical analysis of user expertise on Stack Exchange websites and found that the expertise of users does not increase with time spent in the community; experts join the community as experts, and provide good answers from the beginning~\cite{posnett2012mining}. Vasilescu et al. analyzed how social Q\&A sites such as Stack Exchange affect the knowledge sharing practices in open source communities~\cite{vasilescu2014social}. 
 
\textit{Twitter} also has been explored in recent years by the software engineering research community. Singer et al. did a survey involving 271 developers from GitHub and found that Twitter is used by developers to keep themselves up-to-date with the latest happenings in software development~\cite{singer2014software}. Bougie et al. did an exploratory study on understanding how Twitter is used in software engineering~\cite{Bougie11}. Wang et al. studied the usage of Twitter in Drupal open source development~\cite{Wang13}. Tian et al. found that Twitter is also used by software developers for coordination of efforts, sharing of knowledge, etc.~\cite{Tian14,Tian12does}. Sharma  et al. have explored the categories of software engineering related tweets and events on Twitter~\cite{sharma2015s}.  Methods to filter software-relevant tweets and links have been proposed in Prasetyo et al.~\cite{prasetyo12} and Sharma et al.~\cite{sharma2015nirmal,sharma2017harnessing}.  Sharma et al. also proposed an approach to find software experts on Twitter~\cite{sharma2018recexperts}. Guzman et al. analyzed tweets on Twitter which talked about software applications and companies, and demonstrated that machine learning techniques have the capacity to identify valuable information  for companies and developers of software applications~\cite{guzman2016needle,guzman2017exploratory}. They also proposed a technique to mine tweets for software requirements and evolution~\cite{guzman2017little}.  There has been other work also on mining Twitter feeds for software user requirements such as by William et al.~\cite{williams2017mining}.  Mezouar et al.  found that tweets generated by users can help in early detection of bugs in software applications, and can help developers know about a bug which may be affecting a large user base~\cite{el2018tweets}.

\textit{Software development videos} on YouTube in recent years have been studied as a repository from which software-related knowledge can be extracted. MacLeod et al. studied the  developer's usage of videos (on YouTube) to document software knowledge~\cite{macleod2015code,macleod2017documenting}. They found that the main motivation for sharing videos by developers are building an online identity, to give back to community, to promote themselves, etc.  Ponzanelli et al. proposed an approach to extract relevant fragments from software development video tutorials~\cite{ponzanelli2016codetube,ponzanelli2016too}. Their approach splits the video tutorials into coherent fragments, which are then classified into relevant categories. These fragments are then available individually for developers to query, rather than being forced to browse the whole video. Poche et al. proposed an approach to identify relevant user comments on coding video tutorials on YouTube~\cite{poche2017analyzing}. Parra et al. had proposed a  text-mining-based approach to recommend tags for software development videos on YouTube~\cite{parra2018automatic}. Recently there has been work on extracting code and/or code related features also from programming tutorial videos extracted from YouTube by Yadid et al.~\cite{yadid2016extracting} and Ott et al.~\cite{ott2018learning}.

\subsection{Leveraging Word Embeddings}
Harris et al. had hypothesized that words tend to have similar meaning  in similar contexts~\cite{harris1954distributional}. Mikolov et al. proposed two neural-network-based language models to represent words as a low dimensional vector~\cite{mikolov2013efficient,mikolov2013distributed}. These vectors are commonly known as \textit{word embeddings}. These models have shown considerable success in many NLP tasks~\cite{collobert2011natural,mikolov2013linguistic,baroni2014don}. Word embeddings has been used in software engineering for improving information retrieval tasks~\cite{ye2016word,xu2016domain,chen2016learning}. Chen et al. have used word embeddings based methods to mine analogical libraries~\cite{chen2016mining}, assist collaborative editing~\cite{chen2018edit}, recommend tag synonyms~\cite{chen2017unsupervised} and recommend similar libraries~\cite{chen2016similartech}. In~\cite{yang2016combining}, word embeddings was combined with information retrieval to recommend similar bug reports. Methods similar to or based on word embeddings have also been used recently for better code retrieval~\cite{van2017combining}, to find common software weaknesses~\cite{deepweak}, API recommendation~\cite{zhangrecommending} and sentiment analysis for software engineering~\cite{calefato2017sentiment}.
Our work complements the existing work as we build a cross-platform approach that leverages word embeddings to aid software-development-specific knowledge extraction tasks. Additionally, we demonstrate the value of leveraging word embeddings built from a platform that contains rich software-development-relevant content to solve tasks in another platform.
\section{Conclusion and Future Work}
\label{sec.conclusion}

We proposed an approach to exploit knowledge from rich software-development-specific platforms, to automate knowledge seeking tasks in other less software-development-specific platforms. We first built word embeddings from text extracted from Stack Overflow and Sofware Engineering Stack Exchange, to represent software-development-related knowledge sources. We then leveraged the word embeddings to solve tasks in two different target platforms. In the first use case, we leveraged the word embeddings and sampled sentences from source platforms, to find software-related tweets. In the second use case, we used the word embeddings to classify informative comments on YouTube video tutorials. Based on our experiments conducted in both use cases, our approach improves performance of existing state-of-the-art work for software-development-specific knowledge extraction tasks in the target platforms.

In the future, we intend to perform additional experiments to evaluate the effectiveness of the approach for additional tasks. Finally, we also plan to expand the work to other platforms and knowledge sources, such as Wikipedia articles, software development blogs, README files on GitHub, and software documentation.


\bibliographystyle{plain}      
\bibliography{mainbib}   

%
%

\end{document}